\documentclass{caosp310}

\usepackage{graphicx}

\usepackage{natbib}
\usepackage{comment}
\usepackage{float}
\articleNo{361}
\pubyear{2026}
\volume{54}
\volnumber{3}
\firstpage{1}
\received{May 1, 2024}
\accepted{July 28, 2024}

\def\BibTeX{{\rm B\kern-.05em{\sc i\kern-.025em b}\kern-.08em
             T\kern-.1667em\lower.7ex\hbox{E}\kern-.125emX}}

\begin{document}

\htitle{Exploring Young Stellar Variability with THESEUS}
\hauthor{N.\,Varga, Á.\,Juhász, L. V.\, Tóth}

\title{Exploring Young Stellar Variability with THESEUS: Multi-wavelength Synergies and Gamma-ray Megaflare Prospects}

\author{
        N.\,Varga\inst{1}\orcid{0000-0003-3599-516X}
      \and
        Á.\,Juhász\inst{2} 
      \and
        L. V.\, Tóth\inst{1,2} 
       }

\institute{
           Eötvös Loránd University, Institute of Physics, Pázmány Péter sétány 1/A, H-1117 Budapest, Hungary
         \and 
           University of Debrecen, Institute of Physics, Egyetem tér 1, H-4032 Debrecen, Hungary
          }

\date{March 8, 2026}

\maketitle

\begin{abstract}

The Transient High-Energy Sky and Early Universe Surveyor (THESEUS) will provide new opportunities for time-domain studies of young stellar objects (YSOs) through its soft X-ray, hard X-ray/$\gamma$-ray, and near-infrared capabilities. We investigate the potential of coordinated multi-wavelength observations of Galactic star-forming regions, with emphasis on X-ray flares, ultraviolet follow-up, and the most energetic megaflare events. Since predicted flare yields alone do not fully characterize the suitability of different regions for coordinated observations, we introduce a simple comparative metric that combines flare yield with source distance. The resulting ranking favours nearby star-forming regions for detailed multi-wavelength studies, while distant massive complexes remain valuable for statistical investigations of rare energetic events. We also present an order-of-magnitude estimate of the prospects for detecting gamma-ray megaflares with THESEUS/XGIS based on published \textit{Fermi}-LAT results, illustrating how source distance and stellar population together influence the expected scientific return of high-energy observations.

\keywords{THESEUS -- young stellar objects -- star-forming regions -- multi-wavelength observations -- ultraviolet follow-up -- X-ray flares -- gamma-ray flares}

\end{abstract}
\section{Introduction}
\label{intr}

Young stellar objects (YSOs) are highly variable sources whose emission is driven by magnetic activity, accretion, and circumstellar interactions, producing variability from the X-ray to the near-infrared \citep{Feigelson1999,Audard2014}.

Although THESEUS is primarily designed for gamma-ray burst studies, its Soft X-ray Imager (SXI) and InfraRed Telescope (IRT) also enable investigations of Galactic YSOs through X-ray flares and infrared variability \citep{Amati2018,Amati2021}. Nearby star-forming regions provide representative targets for such observations.

Building on the THESEUS flare-yield estimates published by \citep{Juhasz2026}, we investigate the potential of Galactic star-forming regions for coordinated X-ray, near-infrared, and ultraviolet observations. We extend the previous analysis by introducing a distance-weighted metric to compare the suitability of different regions for multi-wavelength follow-up. In addition, we provide an independent, order-of-magnitude assessment of the prospects for detecting gamma-ray megaflares with THESEUS/XGIS.

\section{THESEUS Instrumentation}

THESEUS carries three instruments covering the soft X-ray, hard X-ray/$\gamma$-ray, and near-infrared wavelength ranges \citep{THESEUS_YellowBook}: the \textit{Soft X-ray Imager} (SXI), the \textit{InfraRed Telescope} (IRT), and the \textit{X-Gamma-ray Imaging Spectrometer} (XGIS).

The SXI operates in the 0.3--5~keV energy range with a field of view of approximately 0.5~sr, enabling simultaneous monitoring of entire nearby star-forming regions and making it well suited for detecting magnetic flares and accretion-related variability in young stellar objects \citep{Feigelson1999,Getman2005b}.

The IRT covers the 0.7--1.8~$\mu$m range and performs imaging and low-resolution spectroscopy in the near-infrared, providing diagnostics of circumstellar material and accretion variability \citep{Audard2014}.

The XGIS extends the energy coverage from 2~keV to 20~MeV over a field of view of approximately $2\pi$~sr, making it suitable for monitoring rare high-energy transients across a large fraction of the sky and motivating the order-of-magnitude estimates presented in Section~\ref{sec:xgis}.

\section{YSO Variability Relevant to THESEUS}

Young stellar objects exhibit variability over timescales ranging from minutes to years owing to magnetic activity, accretion, stellar rotation, and interactions with circumstellar material \citep{Feigelson1999,Audard2014}. These processes produce observable signatures across a broad wavelength range, making YSOs well suited to time-domain studies across multiple wavelengths.

The dominant high-energy signatures are magnetic X-ray flares produced by coronal reconnection events, which form the basis of the flare-rate estimates discussed in this work. In addition, accretion variability can produce changes in both X-ray and near-infrared emission, while variations in circumstellar disks and envelopes give rise to infrared variability and variable X-ray absorption.

Together, these processes provide the physical motivation for coordinated observations with the SXI and IRT instruments. Their complementary multi-wavelength interpretation is discussed in Section~\ref{sec:synergy}.
 
\section{Synergies Between X-ray, Near-Infrared, and Ultraviolet Observations}
\label{sec:synergy}

Young stellar variability produces observable signatures across the X-ray, ultraviolet, and near-infrared domains. Soft X-rays primarily trace magnetic activity, ultraviolet observations probe accretion shocks and flare heating, while near-infrared measurements are sensitive to circumstellar material and accretion-related variability. These complementary diagnostics provide different views of the same physical processes.

The diagnostic value of each wavelength depends on the variability mechanism. Magnetic flares are best constrained by coordinated X-ray and ultraviolet observations, whereas accretion events benefit from simultaneous X-ray, ultraviolet, and near-infrared measurements. Circumstellar variability is primarily investigated through combined X-ray and infrared observations.

\begin{table}[ht]
\centering
\caption{Multi-wavelength diagnostics for YSO variability. P = primary diagnostic, I = important supporting diagnostic, C = complementary diagnostic, -- = limited diagnostic role.}
\label{tab:synergy}
\small
\begin{tabular}{lcccc}
\hline
Diagnostic target & SXI & IRT & UV & Primary scientific objective \\
\hline
Magnetic flares     & P & -- & I & Magnetic activity \\
Accretion bursts    & I & P  & P & Accretion--magnetic coupling \\
Jets \& outflows    & C & C  & C & Shock heating and mass ejection \\
Circumstellar disks & I & P  & -- & Disk structure and evolution \\
Envelope evolution  & I & I  & C & Stellar feedback \\
\hline
Massive regions     & I & C  & C &
Statistical sampling of rare events\\
                    &   &    &   & and massive star formation \\
\hline
\end{tabular}
\end{table}

Table~\ref{tab:synergy} summarizes the relative diagnostic roles of the SXI, IRT, and ultraviolet observations for the principal classes of YSO variability. Although THESEUS has no dedicated ultraviolet instrument, coordinated observations with future UV missions such as ULTRASAT, UVEX, and WSO-UV \citep{Sagiv2014,Kulkarni2021,Shustov2014} can extend its scientific capabilities and provide the observational context for the region comparison presented in the following section.

\section{Representative Star-forming Regions for Multi-wavelength Studies}

Galactic star-forming regions span a wide range of stellar masses, evolutionary stages, and physical environments, providing representative targets for coordinated X-ray, near-infrared, and ultraviolet observations.

\textbf{Benchmark regions:} Orion and Taurus serve as complementary reference environments, supporting statistical studies of stellar populations and detailed investigations of individual YSOs, respectively \citep{Preibisch2005}.

\textbf{Embedded regions:} Complexes such as $\rho$ Ophiuchi, Monoceros~R2, and NGC~6334 probe the earliest stages of star formation, where strong extinction and dense circumstellar material dominate the observed variability \citep{Imanishi2002}.

\textbf{Massive regions:} Carina and Cygnus~X contain large YSO populations, enabling statistical studies of rare energetic events owing to their large source samples, while also providing key laboratories for high-mass star formation \citep{Townsley2011}.

\textbf{Evolutionary sequence:} Nearby associations such as Upper Scorpius extend these studies beyond the main accretion phase, tracing the evolution of magnetic activity in young stars \citep{Preibisch2008}.

These representative regions provide complementary scientific opportunities. Their relative suitability for coordinated ultraviolet follow-up is examined in the following section.

\section{Region-dependent Impact of Ultraviolet Follow-up}

The scientific return of coordinated ultraviolet observations varies among Galactic star-forming regions. While X-ray and near-infrared observations remain effective in a wide range of environments, the usefulness of ultraviolet diagnostics is expected to decrease with increasing distance and extinction.

To compare the potential of coordinated ultraviolet follow-up, we define a simple figure of merit,

\begin{equation}
 {I_{\rm MW}=\frac{\langle N_{\rm flare}\rangle}{d}},
\end{equation}

 where $N_{\rm flare}$ is the predicted number of detectable THESEUS flares reported by \citep{Juhasz2026}, and $d$ is the distance to the corresponding star-forming region. The $1/d$ weighting is adopted as a heuristic proxy for the increasing difficulty of coordinated ultraviolet observations with distance. The proposed metric is not intended as a physical model, but as a convenient comparative measure for prioritizing regions for multi-wavelength follow-up. Alternative weighting schemes could be adopted for different scientific objectives.

\begin{figure}[H]
\centering
\includegraphics[width=\textwidth]{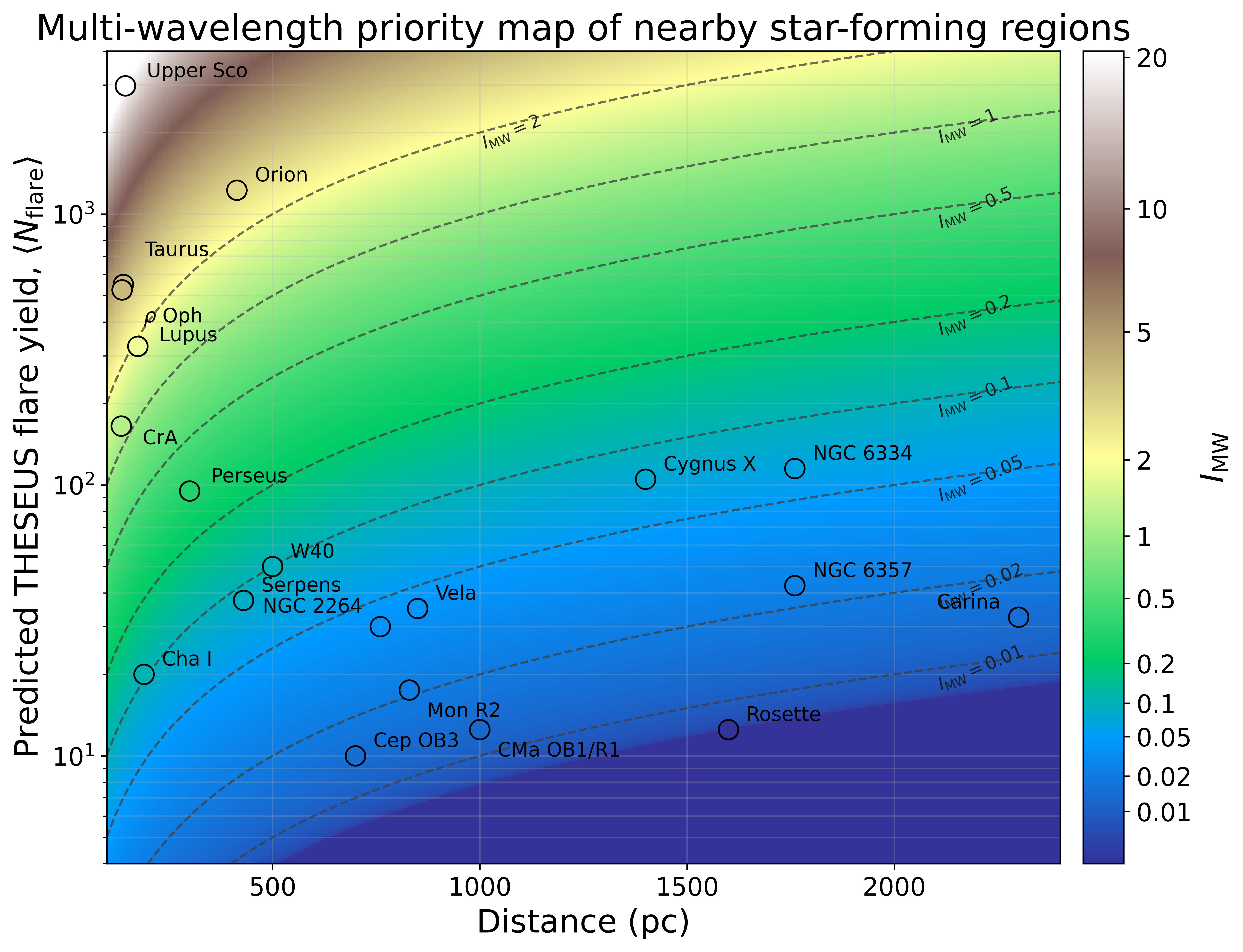}
\caption{Comparison of representative Galactic star-forming regions in the plane of predicted THESEUS flare yield and distance. The coloured background and dashed lines illustrate the analytical behaviour of the adopted comparative metric $I_{\rm MW}=N_{\rm flare}/d$, where $N_{\rm flare}$ represents a general flare-yield value and $d$ is the distance to the region. The plotted regions are positioned using the arithmetic midpoints $\langle N_{\rm flare}\rangle$ of the published four-year flare-yield intervals from \citep{Juhasz2026}. The background is not derived from interpolation of the plotted data points, but represents the continuous distribution of the adopted figure of merit. Regions located toward the upper left combine high predicted flare yields with favourable distances for coordinated ultraviolet follow-up and consequently attain the highest $I_{\rm MW}$ values, whereas distant massive complexes remain important primarily for statistical studies of rare energetic events despite their lower multi-wavelength priority.}
\label{fig:imw}
\end{figure}

Using the flare-yield intervals published by \citep{Juhasz2026}, we calculate a single representative value for each region as the arithmetic midpoint of the corresponding interval, denoted by $\langle N_{\rm flare}\rangle$. The multi-wavelength priority metric is then calculated as $I_{\rm MW}=\langle N_{\rm flare}\rangle/d$, where $d$ is the distance to the region. These midpoint values are used only for comparison and should not be interpreted as independent predictions or precise expected event numbers. The resulting values are summarized in Table~\ref{tab:imw} and illustrated in Fig.~\ref{fig:imw}.
Nearby star-forming regions achieve the highest values of $I_{\rm MW}$, whereas distant massive complexes rank lower despite their much larger stellar populations. This illustrates that the scientific return of coordinated ultraviolet observations depends not only on the expected flare yield but also on the suitability of a region for coordinated ultraviolet follow-up.

\begin{table}[H]
\centering
\caption{Comparison of representative Galactic star-forming regions ranked according to the multi-wavelength priority metric
$I_{\rm MW}=\langle N_{\rm flare}\rangle/d$.
The published four-year flare-yield intervals are taken from
\citep{Juhasz2026}; $\langle N_{\rm flare}\rangle$ is their
arithmetic midpoint. The distances are given in parsecs.}
\label{tab:imw}

\small
\begin{tabular}{lccc}
\hline\hline
Region & $d$ [pc] & $\langle N_{\rm flare}\rangle$ & $I_{\rm MW}$ \\
\hline
Upper Sco      & 145  & 2975.0 & 20.517 \\
Taurus         & 140  & 550.0  & 3.929 \\
$\rho$ Oph     & 137  & 525.0  & 3.832 \\
Orion          & 414  & 1225.0 & 2.959 \\
Lupus          & 175  & 325.0  & 1.857 \\
CrA            & 135  & 165.0  & 1.222 \\
Perseus        & 300  & 95.0   & 0.317 \\
Cha I          & 190  & 20.0   & 0.105 \\
W40            & 500  & 50.0   & 0.100 \\
Serpens        & 430  & 37.5   & 0.087 \\
Cygnus X       & 1400 & 105.0  & 0.075 \\
NGC 6334       & 1760 & 115.0  & 0.065 \\
Vela           & 850  & 35.0   & 0.041 \\
NGC 2264       & 760  & 30.0   & 0.039 \\
NGC 6357       & 1760 & 42.5   & 0.024 \\
Mon R2         & 830  & 17.5   & 0.021 \\
Carina         & 2300 & 32.5   & 0.014 \\
Cep OB3        & 700  & 10.0   & 0.014 \\
CMa OB1/R1     & 1000 & 12.5   & 0.013 \\
Rosette        & 1600 & 12.5   & 0.008 \\
\hline
\end{tabular}
\end{table}

Consequently, nearby regions are expected to provide the most complete multi-wavelength characterization of magnetic activity and accretion, while distant massive complexes remain particularly valuable for statistical studies of rare energetic events.

\section{Prospects for XGIS Observations of YSO Gamma-ray Megaflares}
\label{sec:xgis}

Recent theoretical studies suggest that magnetic megaflares from T Tauri stars may produce high-energy gamma-ray emission, raising the possibility that THESEUS/XGIS could detect the most energetic tail of YSO activity.

Following \citep{Filocomo2023}, we adopt the reported occurrence rate of gamma-ray megaflares as a reference value and assume that it scales approximately with the number of young stellar objects in each star-forming region. For a region containing $N_{\rm YSO}$ young stars, the expected number of megaflares during an effective XGIS observing time $T_{\rm eff}$ is estimated as

\begin{equation}
N_{\rm mega}=N_{\rm YSO}
\left(\frac{\nu_{\rm mega}}{58}\right)
T_{\rm eff},
\end{equation}

where $\nu_{\rm mega}=0.076~{\rm yr^{-1}}$ is the occurrence rate derived by \citep{Filocomo2023} for a sample of 58 T~Tauri stars. Assuming that the very wide XGIS field of view provides an average effective exposure of approximately two years over the nominal four-year THESEUS mission yields the event numbers listed in Table~\ref{tab:xgis}.

To assess detectability, we use the LAT-band energy flux reported by \\ \citep{Filocomo2023} as a reference and scale it with distance assuming an inverse-square law,

\begin{equation}
F_{\rm LAT}(d)=F_0
\left(\frac{d_0}{d}\right)^2,
\end{equation}

where $F_0$ is the published LAT-band energy flux at distance $d_0=390$ pc, with LAT referring to the Large Area Telescope on board the \textit{Fermi} Gamma-ray Space Telescope \citep{Atwood2009}. This scaling is intended only as an order-of-magnitude estimate of the distance dependence and does not imply spectral equivalence between the LAT and XGIS energy bands. We then compare the scaled flux with the nominal XGIS sensitivity reported in the THESEUS Yellow Book \citep{THESEUS_YellowBook} by defining

\begin{equation}
R=
\frac{F_{\rm LAT}(d)}
{F_{\rm XGIS,lim}},
\end{equation}

where $F_{\rm XGIS,lim}$ denotes the 2--30 keV XGIS sensitivity for a representative 10 ks integration. Values of $R>1$ indicate that a comparable flux level, if present in the XGIS band, would exceed the nominal sensitivity, while smaller values imply proportionally higher required X-ray flux.

The results reveal a clear trade-off between source statistics and detectability. Massive complexes such as Carina and NGC~6334 are expected to host the largest numbers of gamma-ray megaflares owing to their very large YSO populations, but their large distances reduce the expected fluxes well below the nominal XGIS sensitivity. Conversely, nearby regions including Upper Scorpius, Taurus, and $\rho$ Ophiuchi exhibit the most favourable detectability despite their smaller expected event numbers. Orion represents an intermediate case, combining a relatively high predicted megaflare rate with favourable detectability, and therefore appears to provide the best compromise between expected megaflare statistics and detectability.

\begin{table}[H]
\centering
\caption{Order-of-magnitude estimate of XGIS prospects for gamma-ray megaflares. $N_{\rm mega}$ is the expected number of Filocomo-like events during an effective two-year XGIS exposure. $R$ is the ratio of the distance-scaled LAT-band flux to the nominal 2--30 keV XGIS sensitivity (10 ks integration).}
\label{tab:xgis}
\small
\begin{tabular}{lccc}
\hline
Region & $N_{\rm mega}$ & $R$ & Comment \\
\hline
Upper Sco & 5.2 & 1.35 & High detectability \\
Taurus    & 1.0 & 1.44 & High detectability \\
$\rho$ Oph& 0.6 & 1.51 & High detectability \\
Orion     & 15.7 & 0.17 & Best overall compromise \\
Cygnus X  & 26.2 & 0.014 & Many events, flux-limited \\
NGC 6334  & 65.5 & 0.009 & Population-rich, flux-limited \\
Carina    & 131  & 0.005 & Largest population, most distant \\
\hline
\end{tabular}
\end{table}

\section{Conclusions}

THESEUS will provide new opportunities for coordinated multi-wavelength studies of young stellar objects through its X-ray, gamma-ray, and near-infrared instrumentation, complemented by future ultraviolet facilities.

To compare the suitability of Galactic star-forming regions for coordinated observations, we introduce the simple comparative metric $I_{\rm MW}$, which combines the predicted THESEUS flare yield with source distance. The resulting ranking (Table~\ref{tab:imw}, Fig.~\ref{fig:imw}) favours nearby regions for detailed multi-wavelength studies, while distant massive complexes remain particularly valuable for statistical investigations of rare energetic events and high-mass star formation.

We also present an order-of-magnitude assessment of XGIS prospects for detecting gamma-ray megaflares using published \textit{Fermi}-LAT predictions. The corresponding parameter $R$, defined as the ratio of the distance-scaled LAT flux to the nominal XGIS sensitivity, indicates that nearby regions provide the most favourable conditions for individual detections, whereas distant massive complexes are expected to contribute primarily through their large YSO populations.

Together, these complementary metrics provide a practical framework for prioritizing Galactic star-forming regions for future coordinated THESEUS observations across the X-ray, gamma-ray, near-infrared, and ultraviolet domains.

\bibliography{demo_caosp310}

\begin{thebibliography}{16}
\expandafter\ifx\csname natexlab\endcsname\relax\def\natexlab#1{#1}\fi

%ADS_ID Amati2018
\bibitem[{{Amati} {et~al.}(2018){Amati}, {O'Brien}, {G{\"o}tz}, {Bozzo},
  {Tenzer}, {Frontera}, {Ghirlanda}, {Labanti}, {Osborne}, {Stratta}, {Tanvir},
  {Willingale}, {Attina}, {Campana}, {Castro-Tirado}, {Contini}, {Fuschino},
  {Gomboc}, {Hudec}, {Orleanski}, {Renotte}, {Rodic}, {Bagoly}, {Blain},
  {Callanan}, {Covino}, {Ferrara}, {Le Floch}, {Marisaldi}, {Mereghetti},
  {Rosati}, {Vacchi}, {D'Avanzo}, {Giommi}, {Piranomonte}, {Piro}, {Reglero},
  {Rossi}, {Santangelo}, {Salvaterra}, {Tagliaferri}, {Vergani}, {Vinciguerra},
  {Briggs}, {Campolongo}, {Ciolfi}, {Connaughton}, {Cordier}, {Morelli},
  {Orlandini}, {Adami}, {Argan}, {Atteia}, {Auricchio}, {Balazs}, {Baldazzi},
  {Basa}, {Basak}, {Bellutti}, {Bernardini}, {Bertuccio}, {Braga}, {Branchesi},
  {Brandt}, {Brocato}, {Budtz-Jorgensen}, {Bulgarelli}, {Burderi}, {Camp},
  {Capozziello}, {Caruana}, {Casella}, {Cenko}, {Chardonnet}, {Ciardi},
  {Colafrancesco}, {Dainotti}, {D'Elia}, {De Martino}, {De Pasquale}, {Del
  Monte}, {Della Valle}, {Drago}, {Evangelista}, {Feroci}, {Finelli},
  {Fiorini}, {Fynbo}, {Gal-Yam}, {Gendre}, {Ghisellini}, {Grado}, {Guidorzi},
  {Hafizi}, {Hanlon}, {Hjorth}, {Izzo}, {Kiss}, {Kumar}, {Kuvvetli}, {Lavagna},
  {Li}, {Longo}, {Lyutikov}, {Maio}, {Maiorano}, {Malcovati}, {Malesani},
  {Margutti}, {Martin-Carrillo}, {Masetti}, {McBreen}, {Mignani}, {Morgante},
  {Mundell}, {Nargaard-Nielsen}, {Nicastro}, {Palazzi}, {Paltani}, {Panessa},
  {Pareschi}, {Pe'er}, {Penacchioni}, {Pian}, {Piedipalumbo}, {Piran}, {Rauw},
  {Razzano}, {Read}, {Rezzolla}, {Romano}, {Ruffini}, {Savaglio}, {Sguera},
  {Schady}, {Skidmore}, {Song}, {Stanway}, {Starling}, {Topinka}, {Troja}, {van
  Putten}, {Vanzella}, {Vercellone}, {Wilson-Hodge}, {Yonetoku}, {Zampa},
  {Zampa}, {Zhang}, {Zhang}, {Zhang}, {Zhang}, {Antonelli}, {Bianco}, {Boci},
  {Boer}, {Botticella}, {Boulade}, {Butler}, {Campana}, {Capitanio}, {Celotti},
  {Chen}, {Colpi}, {Comastri}, {Cuby}, {Dadina}, {De Luca}, {Dong}, {Ettori},
  {Gandhi}, {Geza}, {Greiner}, {Guiriec}, {Harms}, {Hernanz}, {Hornstrup},
  {Hutchinson}, {Israel}, {Jonker}, {Kaneko}, {Kawai}, {Wiersema}, {Korpela},
  {Lebrun}, {Lu}, {MacFadyen}, {Malaguti}, {Maraschi}, {Melandri}, {Modjaz},
  {Morris}, {Omodei}, {Paizis}, {P{\'a}ta}, {Petrosian}, {Rachevski}, {Rhoads},
  {Ryde}, \& {Sabau-Graziati}}]{Amati2018}
{Amati}, L., {O'Brien}, P., {G{\"o}tz}, D., {et~al.}, {The THESEUS space
  mission concept: science case, design and expected performances}. 2018, {\it
  Advances in Space Research}, {\bf 62}, 191,
  \href{https://doi.org/10.1016/j.asr.2018.03.010}{\path{DOI:
  10.1016/j.asr.2018.03.010}}

%ADS_ID Amati2021
\bibitem[{{Amati} {et~al.}(2021){Amati}, {O'Brien}, {G{\"o}tz}, {Bozzo},
  {Santangelo}, {Tanvir}, {Frontera}, {Mereghetti}, {Osborne}, {Blain}, {Basa},
  {Branchesi}, {Burderi}, {Caballero-Garc{\'\i}a}, {Castro-Tirado},
  {Christensen}, {Ciolfi}, {De Rosa}, {Doroshenko}, {Ferrara}, {Ghirlanda},
  {Hanlon}, {Heddermann}, {Hutchinson}, {Labanti}, {Le Floch}, {Lerman},
  {Paltani}, {Reglero}, {Rezzolla}, {Rosati}, {Salvaterra}, {Stratta},
  {Tenzer}, \& {Theseus Consortium}}]{Amati2021}
{Amati}, L., {O'Brien}, P.~T., {G{\"o}tz}, D., {et~al.}, {The THESEUS space
  mission: science goals, requirements and mission concept}. 2021, {\it
  Experimental Astronomy}, {\bf 52}, 183,
  \href{https://doi.org/10.1007/s10686-021-09807-8}{\path{DOI:
  10.1007/s10686-021-09807-8}}

%ADS_ID Atwood2009
\bibitem[{{Atwood} {et~al.}(2009){Atwood}, {Abdo}, {Ackermann}, {Althouse},
  {Anderson}, {Axelsson}, {Baldini}, {Ballet}, {Band}, {Barbiellini},
  {Bartelt}, {Bastieri}, {Baughman}, {Bechtol}, {B{\'e}d{\'e}r{\`e}de},
  {Bellardi}, {Bellazzini}, {Berenji}, {Bignami}, {Bisello}, {Bissaldi},
  {Blandford}, {Bloom}, {Bogart}, {Bonamente}, {Bonnell}, {Borgland},
  {Bouvier}, {Bregeon}, {Brez}, {Brigida}, {Bruel}, {Burnett}, {Busetto},
  {Caliandro}, {Cameron}, {Caraveo}, {Carius}, {Carlson}, {Casandjian},
  {Cavazzuti}, {Ceccanti}, {Cecchi}, {Charles}, {Chekhtman}, {Cheung},
  {Chiang}, {Chipaux}, {Cillis}, {Ciprini}, {Claus}, {Cohen-Tanugi},
  {Condamoor}, {Conrad}, {Corbet}, {Corucci}, {Costamante}, {Cutini}, {Davis},
  {Decotigny}, {DeKlotz}, {Dermer}, {de Angelis}, {Digel}, {do Couto e Silva},
  {Drell}, {Dubois}, {Dumora}, {Edmonds}, {Fabiani}, {Farnier}, {Favuzzi},
  {Flath}, {Fleury}, {Focke}, {Funk}, {Fusco}, {Gargano}, {Gasparrini},
  {Gehrels}, {Gentit}, {Germani}, {Giebels}, {Giglietto}, {Giommi}, {Giordano},
  {Glanzman}, {Godfrey}, {Grenier}, {Grondin}, {Grove}, {Guillemot}, {Guiriec},
  {Haller}, {Harding}, {Hart}, {Hays}, {Healey}, {Hirayama}, {Hjalmarsdotter},
  {Horn}, {Hughes}, {J{\'o}hannesson}, {Johansson}, {Johnson}, {Johnson},
  {Johnson}, {Johnson}, {Kamae}, {Katagiri}, {Kataoka}, {Kavelaars}, {Kawai},
  {Kelly}, {Kerr}, {Klamra}, {Kn{\"o}dlseder}, {Kocian}, {Komin}, {Kuehn},
  {Kuss}, {Landriu}, {Latronico}, {Lee}, {Lee}, {Lemoine-Goumard}, {Lionetto},
  {Longo}, {Loparco}, {Lott}, {Lovellette}, {Lubrano}, {Madejski}, {Makeev},
  {Marangelli}, {Massai}, {Mazziotta}, {McEnery}, {Menon}, {Meurer},
  {Michelson}, {Minuti}, {Mirizzi}, {Mitthumsiri}, {Mizuno}, {Moiseev},
  {Monte}, {Monzani}, {Moretti}, {Morselli}, {Moskalenko}, {Murgia},
  {Nakamori}, {Nishino}, {Nolan}, {Norris}, {Nuss}, {Ohno}, {Ohsugi}, {Omodei},
  {Orlando}, {Ormes}, {Paccagnella}, {Paneque}, {Panetta}, {Parent}, {Pearce},
  {Pepe}, {Perazzo}, {Pesce-Rollins}, {Picozza}, {Pieri}, {Pinchera}, {Piron},
  {Porter}, {Poupard}, {Rain{\`o}}, {Rando}, {Rapposelli}, {Razzano}, {Reimer},
  {Reimer}, {Reposeur}, {Reyes}, {Ritz}, {Rochester}, {Rodriguez}, {Romani},
  {Roth}, {Russell}, {Ryde}, {Sabatini}, {Sadrozinski}, {Sanchez}, {Sander},
  {Sapozhnikov}, {Parkinson}, {Scargle}, {Schalk}, \& {Scolieri}}]{Atwood2009}
{Atwood}, W.~B., {Abdo}, A.~A., {Ackermann}, M., {et~al.}, {The Large Area
  Telescope on the Fermi Gamma-Ray Space Telescope Mission}. 2009, {\it \apj},
  {\bf 697}, 1071,
  \href{https://doi.org/10.1088/0004-637X/697/2/1071}{\path{DOI:
  10.1088/0004-637X/697/2/1071}}

%ADS_ID Audard2014
\bibitem[{{Audard} {et~al.}(2014){Audard}, {{\'A}brah{\'a}m}, {Dunham},
  {Green}, {Grosso}, {Hamaguchi}, {Kastner}, {K{\'o}sp{\'a}l}, {Lodato},
  {Romanova}, {Skinner}, {Vorobyov}, \& {Zhu}}]{Audard2014}
{Audard}, M., {{\'A}brah{\'a}m}, P., {Dunham}, M.~M., {et~al.}, {Episodic
  Accretion in Young Stars}. 2014, in {\it Protostars and Planets VI}, ed.
  H.~{Beuther}, R.~S. {Klessen}, C.~P. {Dullemond}, \& T.~{Henning}, 387--410

%ADS_ID THESEUS_YellowBook
\bibitem[{{ESA/THESEUS Study Team}(2026)}]{THESEUS_YellowBook}
{ESA/THESEUS Study Team}. 2026, {THESEUS: Transient High-Energy Sky and Early
  Universe Surveyor Assessment Study Report}, Assessment study report, European
  Space Agency, accessed: 2026-06-30

%ADS_ID Feigelson1999
\bibitem[{{Feigelson} \& {Montmerle}(1999)}]{Feigelson1999}
{Feigelson}, E.~D. \& {Montmerle}, T., {High-Energy Processes in Young Stellar
  Objects}. 1999, {\it \araa}, {\bf 37}, 363,
  \href{https://doi.org/10.1146/annurev.astro.37.1.363}{\path{DOI:
  10.1146/annurev.astro.37.1.363}}

%ADS_ID Filocomo2023
\bibitem[{{Fil{\'o}como} {et~al.}(2023){Fil{\'o}como}, {Albacete-Colombo},
  {Mestre}, {Pellizza}, \& {Combi}}]{Filocomo2023}
{Fil{\'o}como}, A., {Albacete-Colombo}, J.~F., {Mestre}, E., {Pellizza}, L.~J.,
  \& {Combi}, J.~A., {{\ensuremath{\gamma}}-ray detection from occasional
  flares in T Tauri stars of NGC 2071 - I. Observational connection}. 2023,
  {\it \mnras}, {\bf 525}, 1726,
  \href{https://doi.org/10.1093/mnras/stad2029}{\path{DOI:
  10.1093/mnras/stad2029}}

%ADS_ID Getman2005b
\bibitem[{{Getman} {et~al.}(2005){Getman}, {Flaccomio}, {Broos}, {Grosso},
  {Tsujimoto}, {Townsley}, {Garmire}, {Kastner}, {Li}, {Harnden}, {Wolk},
  {Murray}, {Lada}, {Muench}, {McCaughrean}, {Meeus}, {Damiani}, {Micela},
  {Sciortino}, {Bally}, {Hillenbrand}, {Herbst}, {Preibisch}, \&
  {Feigelson}}]{Getman2005b}
{Getman}, K.~V., {Flaccomio}, E., {Broos}, P.~S., {et~al.}, {Chandra Orion
  Ultradeep Project: Observations and Source Lists}. 2005, {\it \apjs}, {\bf
  160}, 319, \href{https://doi.org/10.1086/432092}{\path{DOI: 10.1086/432092}}

%ADS_ID Imanishi2002
\bibitem[{Imanishi {et~al.}(2002)Imanishi, Koyama, \& Tsuboi}]{Imanishi2002}
Imanishi, K., Koyama, K., \& Tsuboi, Y., Erratum: “Chandra Observation of the
  {$\rho$} Ophiuchi Cloud” (ApJ, 557, 747 [2001]). 2002, {\it The
  Astrophysical Journal}, {\bf 579}, 920,
  \href{https://doi.org/10.1086/342987}{\path{DOI: 10.1086/342987}}

%ADS_ID Juhasz2026
\bibitem[{{Juh{\'a}sz} {et~al.}(2026){Juh{\'a}sz}, {Varga}, {{\'A}brah{\'a}m},
  {Amati}, {Blain}, {Bozzo}, {G{\"u}del}, {G{\"o}tz}, {Massi}, {Meuris},
  {O'Brien}, {Santangelo}, {Vavrek}, \& {T{\'o}th}}]{Juhasz2026}
{Juh{\'a}sz}, {\'A}., {Varga}, N., {{\'A}brah{\'a}m}, P., {et~al.}, {The
  Potential of the THESEUS Mission for Studies of Young Stellar Objects}. 2026,
  {\it Particles}, {\bf 9}, 91,
  \href{https://doi.org/10.3390/particles9030091}{\path{DOI:
  10.3390/particles9030091}}

%ADS_ID Kulkarni2021
\bibitem[{{Kulkarni} {et~al.}(2021){Kulkarni}, {Harrison}, {Grefenstette},
  {Earnshaw}, {Andreoni}, {Berg}, {Bloom}, {Cenko}, {Chornock}, {Christiansen},
  {Coughlin}, {Wuollet Criswell}, {Darvish}, {Das}, {De}, {Dessart}, {Dixon},
  {Dorsman}, {El-Badry}, {Evans}, {Ford}, {Fremling}, {Gansicke}, {Gezari},
  {Goetberg}, {Green}, {Graham}, {Heida}, {Ho}, {Jaodand}, {Johns-Krull},
  {Kasliwal}, {Lazzarini}, {Lu}, {Margutti}, {Martin}, {Masters}, {McKernan},
  {Naze}, {Nissanke}, {Parazin}, {Perley}, {Phinney}, {Piro}, {Raaijmakers},
  {Rauw}, {Rodriguez}, {Sana}, {Senchyna}, {Singer}, {Spake}, {Stassun},
  {Stern}, {Teplitz}, {Weisz}, \& {Yao}}]{Kulkarni2021}
{Kulkarni}, S.~R., {Harrison}, F.~A., {Grefenstette}, B.~W., {et~al.}, {Science
  with the Ultraviolet Explorer (UVEX)}. 2021, {\it arXiv e-prints},
  arXiv:2111.15608, \href{https://doi.org/10.48550/arXiv.2111.15608}{\path{DOI:
  10.48550/arXiv.2111.15608}}

%ADS_ID Preibisch2005
\bibitem[{{Preibisch} {et~al.}(2005){Preibisch}, {Kim}, {Favata}, {Feigelson},
  {Flaccomio}, {Getman}, {Micela}, {Sciortino}, {Stassun}, {Stelzer}, \&
  {Zinnecker}}]{Preibisch2005}
{Preibisch}, T., {Kim}, Y.-C., {Favata}, F., {et~al.}, {The Origin of T Tauri
  X-Ray Emission: New Insights from the Chandra Orion Ultradeep Project}. 2005,
  {\it \apjs}, {\bf 160}, 401, \href{https://doi.org/10.1086/432891}{\path{DOI:
  10.1086/432891}}

%ADS_ID Preibisch2008
\bibitem[{{Preibisch} \& {Mamajek}(2008)}]{Preibisch2008}
{Preibisch}, T. \& {Mamajek}, E., {The Nearest OB Association:
  Scorpius-Centaurus (Sco OB2)}. in , {\it Handbook of Star Forming Regions,
  Volume II}, ed. B.~{Reipurth}, Vol. {\bf ~5} (Astronomical Society of the
  Pacific), 235

%ADS_ID Sagiv2014
\bibitem[{{Sagiv} {et~al.}(2014){Sagiv}, {Gal-Yam}, {Ofek}, {Waxman},
  {Aharonson}, {Kulkarni}, {Nakar}, {Maoz}, {Trakhtenbrot}, {Phinney}, {Topaz},
  {Beichman}, {Murthy}, \& {Worden}}]{Sagiv2014}
{Sagiv}, I., {Gal-Yam}, A., {Ofek}, E.~O., {et~al.}, {Science with a Wide-field
  UV Transient Explorer}. 2014, {\it \aj}, {\bf 147}, 79,
  \href{https://doi.org/10.1088/0004-6256/147/4/79}{\path{DOI:
  10.1088/0004-6256/147/4/79}}

%ADS_ID Shustov2014
\bibitem[{{Shustov} {et~al.}(2014){Shustov}, {G{\'o}mez de Castro}, {Sachkov},
  {Moisheev}, {Kanev}, {L{\'o}pez-Santiago}, {Malkov}, {Nasonov}, {Bel{\'e}n
  Perea}, {S{\'a}nchez}, {Savanov}, {Shugarov}, {Sichevskiy}, {Vlasenko}, \&
  {Ya{\~n}ez}}]{Shustov2014}
{Shustov}, B., {G{\'o}mez de Castro}, A.~I., {Sachkov}, M., {et~al.}, {WSO-UV
  progress and expectations}. 2014, {\it \apss}, {\bf 354}, 155,
  \href{https://doi.org/10.1007/s10509-014-2119-0}{\path{DOI:
  10.1007/s10509-014-2119-0}}

%ADS_ID Townsley2011
\bibitem[{{Townsley} {et~al.}(2011){Townsley}, {Broos}, {Corcoran},
  {Feigelson}, {Gagn{\'e}}, {Montmerle}, {Oey}, {Smith}, {Garmire}, {Getman},
  {Povich}, {Remage Evans}, {Naz{\'e}}, {Parkin}, {Preibisch}, {Wang}, {Wolk},
  {Chu}, {Cohen}, {Gruendl}, {Hamaguchi}, {King}, {Mac Low}, {McCaughrean},
  {Moffat}, {Oskinova}, {Pittard}, {Stassun}, {ud-Doula}, {Walborn}, {Waldron},
  {Churchwell}, {Nichols}, {Owocki}, \& {Schulz}}]{Townsley2011}
{Townsley}, L.~K., {Broos}, P.~S., {Corcoran}, M.~F., {et~al.}, {An
  Introduction to the Chandra Carina Complex Project}. 2011, {\it \apjs}, {\bf
  194}, 1, \href{https://doi.org/10.1088/0067-0049/194/1/1}{\path{DOI:
  10.1088/0067-0049/194/1/1}}

\end{thebibliography}

\end{document}